\documentclass{timgad}
\usepackage{amssymb} 
\usepackage[english]{babel}
\usepackage{amsmath}
\usepackage{amsfonts}
\usepackage{amsthm}
\usepackage{url}
\usepackage{hyperref}
\usepackage{textcomp}
\usepackage{float}
\usepackage{subfig}
\usepackage{nopageno}
\usepackage{enumitem}
\usepackage{tikz}
\usepackage{xspace}
\usepackage{comment}
\usepackage{bm}
\usepackage{MnSymbol}
\usetikzlibrary{backgrounds,plothandlers,plotmarks,cd}

\usepackage[backend=bibtex,bibencoding=ascii,style=alphabetic,bibstyle=alphabetic,firstinits=true,doi=false,isbn=false,url=false,maxbibnames=99]{biblatex}
\renewbibmacro{in:}{}
\AtEveryBibitem{\clearname{editor}}

\newbibmacro{string+link}[1]{%
	\iffieldundef{ee}
	{\iffieldundef{url}
		{\iffieldundef{doi}
			{\iffieldundef{eprint}
				{#1}
				{\href{http://arxiv.org/abs/\thefield{eprint}}{#1}}}
			{\href{http://dx.doi.org/\thefield{doi}}{#1}}}
		{\href{\thefield{url}}{#1}}}
	{\href{\thefield{ee}}{#1}}} 
\DeclareFieldFormat{title}{\usebibmacro{string+link}{\mkbibemph{#1}}}
\DeclareFieldFormat[article,inproceedings,inbook,incollection,mastersthesis,thesis]{title}{\usebibmacro{string+link}{\mkbibquote{#1}}}

\theoremstyle{bgnumberbefore}
\newtheorem{theorem2}{Theorem}[]
\newtheorem{lemma2}[theorem2]{Lemma}

\newcommand{\theTitle}{A note on hardness of promise hypergraph colouring}

\title{\theTitle}
\author{Marcin Wrochna}
\affil{University of Warsaw (Institute of Informatics), Poland}
\date{(written July 2020, sent in private communication, first published online May 2022)}

\begin{document}


\maketitle

\def\HH{\mathbf{H}}
\def\PCSP{\operatorname{PCSP}}
\def\NAE{\operatorname{NAE}}
\def\KG{\operatorname{KG}}
\def\KK{K}
\def\supp{\operatorname{supp}}
\def\sel{\operatorname{sel}}

The goal of this short note is to show a slightly simpler proof of the following.
\begin{theorem2}[Dinur, Regev, Smith~\cite{DRS05}]\label{thm:main}
	$\PCSP(\HH_k, \HH_c)$ is NP-hard for all $2 \leq k\leq c$.
\end{theorem2}

That is, hardness of Promise Hypergraph Colouring (for $2 \leq k\leq c$ and 3-uniform hypergraphs). 
Here $\HH_c$ is the structure with domain $[c]$ and one 3-ary relation $\operatorname{NAE}_c = [c]^3 \setminus \{(a,a,a) \mid a \in [c]\}$.
The problem is: given a 3-uniform hypergraph, can we distinguish the case where it is $k$-colourable (admits homomorphism to $\HH_k$) from the case where it is not even $c$-colourable (does not admit a homomorphism to $\HH_c$; the \emph{promise} is that the input falls into one of these two cases).
The same proof applies to the search version: given a $k$-colourable 3-uniform hypergraph, find a $c$-colouring.

The proof in~\cite{DRS05} relies on a constructing a somewhat ad-hoc reduction and analysing it's completeness and soundness.
We recast this proof in the more recent algebraic framework for Promise CSPs~\cite{BartoBKO21}, where it suffices to study polymorphisms associated with the problem (defined below) and then apply general theorems that give generic NP-hardness reductions.
Thus instead of a more quantitative analysis, we use only constants everywhere (independent of the arity $L$ of the multivariate functions involved).%
\footnote{On the other hand, while the proof in~\cite{DRS05} can be slightly strengthened to statements about independent sets or non-constant numbers of colours, the algebraic framework so far cannot do that directly.}
We also replace the use of Schrijver graphs with the simpler Kneser graphs $\KG(n,k)$, plus a (correct and very easy) case of Hedetniemi's conjecture.
Most importantly, this allows us to derive the result from ``baby'' versions of the PCP theorem, which were recently given purely combinatorial proofs by Barto and Kozik~\cite{BartoK22}, as discussed below.

The two facts we need about colourings are the following.
\begin{theorem2}[Lov\'{a}sz~\cite{Lovasz78}]\label{thm:kneser}
	$\chi(\KG(n,k)) = n - 2k + 2$.
\end{theorem2}

\begin{lemma2}
	Let $\chi(G) > n$. Then $\chi(G \times \KK_{n+1}) > n$.
\end{lemma2}
\begin{proof}
	We show the contrapositive: suppose there is a homomorphism $G \times \KK_{n+1} \to \KK_n$.
	Equivalently $G \to \KK_n^{\KK_{n+1}}$.
	There is a homomorphism $\KK_n^{\KK_{n+1}} \to \KK_n$: simply map any (improper) $n$-colouring of $\KK_{n+1}$ to an arbitrary repeating colour.
	This shows $G \to \KK_n$.
\end{proof}

\noindent
(Alternatively one could say both Kneser graphs and cliques have large chromatic number for topological reasons, so their tensor product does as well~\cite{SimonyiZ10}).

\medskip
To begin the proof of Theorem~\ref{thm:main}, without loss of generality, let $k=2$, $c \geq 2$.
A polymorphism of arity $L$ is a function $f \colon [2]^L \to [c]$ such that for any three input row vectors satisfying $\NAE_2$ column-wise, the outputs are in $\NAE_c$.
Let us consider an arbitrary such polymorphism $f$.
We aim to show that $f$ in a way ``distinguishes'' a constant number of coordinates in $[L]$.

For a set of colours $C \subseteq [c]$, we say a set $S \subseteq [L]$ is \emph{$C$-avoiding} if fixing the inputs on $S$ of $f$ to~$1$ avoids (at least) $C$ in the output.
That is, for every input $\bar{v} \in [2]^L$ with $\bar{v}|_S \equiv 1$ we have $f(\bar{v}) \not\in C$.
A set $S$ is \emph{$t$-avoiding} if it is $C$-avoiding for some set $C$ of size $t$.
One of the main ideas of~\cite{DRS05} is the following lemma.

\begin{lemma2}\label{lem:lem}
	There is a 1-avoiding set $S$ of constant size, $|S| \leq c$.
\end{lemma2}
\begin{proof}
	If  $L \leq c$, then $[L]$ is a $(c-1)$-avoiding set and we are done.
	Otherwise assume $L > c$.
	
	Let $h$ be an integer satisfying $L - c + 2 > 2h \geq L-c$.
	Consider inputs in $[2]^L$ of Hamming~weight~$h$.
	The graph whose vertices are those inputs and whose edges are input pairs with disjoint supports is a Kneser graph $\KG(L,h)$.
	Since $\chi(\KG(L,h))=L-2h+2 > c$,
	there are two inputs $\bar{u},\bar{v}$ with disjoint supports but the same colour $f(\bar{u})=f(\bar{v}) = b$.
	Let $S := [L] \setminus \left(\supp(\bar{u}) \cup \supp(\bar{v})\right)$.
	Then any input $\bar{w}$ with $\bar{w}|_S \equiv 1$ has $f(\bar{w}) \neq b$.
	Thus $S$ is 1-avoiding and $|S| = L - 2h \leq c$.
\end{proof}

The problem with applying this directly is that there may be many disjoint 1-avoiding sets.
However in that case, there are many disjoint sets that avoid the same colour $b$, and we can use a similar argument to find two inputs with the same colour $b' \neq b$ and with large disjoint supports, which similarly implies a small set avoiding $\{b,b'\}$.
More generally the inductive step is as follows.

\begin{lemma2}
	Let $1\leq t < c$ and $L$ large enough ($L \geq (c+1) c^t + c$).
	Suppose $f$ has $>\binom{c}{t}\cdot c$ disjoint $t$-avoiding sets of size $\leq c^t$.
	Then it has a $(t+1)$-avoiding set of size $\leq c^{t+1}$.	
\end{lemma2}
\begin{proof}
	By assumption, $f$ has $\geq c+1$ disjoint sets $S_1,\dots,S_{c+1}\subseteq [L]$ that avoid the same $C \subseteq [c]$ of size $|C|=t$.
	Let $R := [L] \setminus (S_1 \cup \dots \cup S_{c+1})$.
	Let $h$ be an integer satisfying $|R|-c+2 > 2h \geq |R|-c$.
	Consider inputs in  $[2]^L$
	whose support consists of exactly one of $S_1,\dots,S_{c+1}$, plus exactly $h$ elements of $R$.
	The graph whose vertices are those inputs and whose edges are input pairs with disjoint supports is isomorphic to $\KK_{c+1} \times \KG(R,h)$.
	Since $\chi(\KK_{c+1} \times \KG(R,h)) > c$,
	there are two inputs $\bar{u},\bar{v}$ with disjoint support but the same colour $f(\bar{u})=f(\bar{v}) = b'$.
	Since fixing any $S_i$ to 1 avoids $C$, we have $b' \not\in C$.
	Let $S := [L] \setminus \left(\supp(\bar{u}) \cup \supp(\bar{v})\right)$. 
	Then any input $\bar{w}$ with $\bar{w}|_S \equiv 1$ has $f(\bar{w}) \neq b'$.
	Moreover, $\bar{w}|_S \equiv 1$ implies  $\bar{w}|_{S_i} \equiv 1$ for $(c+1)-2 \geq 1$ different $i$, hence $f(\bar{w}) \not \in C$.
	Thus $S$ is $C \cup \{b'\}$-avoiding.
	Finally $|S| \leq ((c+1)-2) \cdot c^t + |R|-2h \leq (c-1) \cdot c^t + c \leq c^{t+1}$.
\end{proof}

\bigskip 

Note that there cannot be any $c$-avoiding set, since $f$ cannot avoid all colours.
So there is a maximum $t=t(f)$ with $1 \leq t <c$ such that $f$ has some $t$-avoiding set of size $\leq c^t$.
Let $\sel(f) \subseteq [L]$ be the sum of a maximal family of disjoint $t(f)$-avoiding sets of size $\leq c^{t(f)}$.
By the above lemma $|\sel(f)| \leq \binom{c}{t} \cdot c \cdot c^t \leq (2c)^c$ (technically if $L < (c+1)c^t+c$ take $\sel(f) = [L]$ instead).

We show that our selection $\sel(f) \subseteq[L]$ of constant size ($\leq (2c)^c$, independent of $L$) is somewhat consistent for different $f$, in the following very weak sense.
Let $g$ be a minor of $f$ for some $\pi\colon [L] \to [L']$;
that is, $g(x_1,\dots,x_{L'}) = f(x_{\pi(1)}, \dots, x_{\pi(L)})$.
We denote this as $f \xrightarrow{\pi} g$.
Suppose for a moment that $t(f)=t(g)=t$.
Clearly for every $t$-avoiding set $S$ of $f$, $\pi(S)$ is a $t$-avoiding set of $g$ of at most the same size.
Hence $\pi(\sel(f))$ intersects $\sel(g)$, by maximality of the family selected for $g$.

Consider now a chain of minors $f_0 \xrightarrow{\pi_{0,1}} f_1 \xrightarrow{\pi_{1,2}} f_2 \dots \xrightarrow{\pi_{c-1, c}} f_c$.
For $i,j$ we let $\pi_{i,j}$ be the composition $\pi_{j-1, j} \circ \dots \circ \pi_{i,i+1}$ and observe that $f_i \xrightarrow{\pi_{i,j}} f_j$.
Since there are only $c-1$ possibilities for what $t(f)$ can be,
we conclude that in any such sequence of minors, there exist $i,j$ such that $t(f_i)=t(f_j)$, and hence $\pi_{i,j}(\sel(f_i))$ intersects $\sel(f_j)$.

Thus we have found a selection $\sel(f) \subseteq[L]$ of constant size for each polymorphism $f$, which is ``consistent'' on chains of minors of length $c$.
This is a sufficient condition for NP-hardness by the following, ``layered, baby'' corollary of the PCP theorem.

\begin{theorem2}[\cite{BrandtsWZ21}]\label{thm:pcp}
	Suppose there are constants $k,\ell$ and there is an assignment $\sel(f) \subseteq[L]$ to every polymorphism $f$ of a Promise CSP, such that for every $\ell$-chain of minors there are $i,j$ such that $\pi_{i,j}(\sel(f_i)) \cap \sel(f_j) \neq \emptyset$. Then the Promise CSP is NP-hard.
\end{theorem2}

(The version in Theorem 5.22 of~\cite{BartoBKO21} would also be sufficient.)
This concludes the proof of Theorem~\ref{thm:main}.

\subsubsection*{Discussion}
Barto and Kozik~\cite{BartoK22} recently found a purely combinatorial, self-contained proof of Theorem~\ref{thm:pcp},
without using the original PCP theorem itself.
The fact that it implies Theorem~\ref{thm:main} (among now many other examples) highlights that the ``baby'' version, despite its humble name, is quite powerful,
making its new proof all the more intriguing.

We remark that the idea of using ``layered'' versions of PCP (here they appear as ``chains of minors'') comes from~\cite{DRS05}.
While it is a relatively simple extension of the non-layered versions, it is now a crucial ingredient of several NP-hardness results, such as in~\cite{BrandtsWZ21}.
We refer the reader to~\cite{BartoK22} for a more detailed discussion of all these versions of the PCP theorem and its corollaries.

On a final note, we stress that at the core of our proof is still Lov\'{a}sz' topological proof of Theorem~\ref{thm:kneser} (relying on the Borsuk-Ulam theorem), as used in Lemma~\ref{lem:lem}.
These ideas from~\cite{DRS05} were extended in other results on Promise CSPs, see e.g.~\cite{AustrinBP20}.
An interesting open problem is whether they can be connected with the other use of topology in proving hardness of Promise CSPs, presented in~\cite{KOWZ}.

\printbibliography

@inproceedings{BartoK22,
  author    = {Libor Barto and
               Marcin Kozik},
  editor    = {Joseph (Seffi) Naor and
               Niv Buchbinder},
  title     = {Combinatorial Gap Theorem and Reductions between Promise CSPs},
  booktitle = {Proceedings of the 2022 {ACM-SIAM} Symposium on Discrete Algorithms,
               {SODA} 2022, Virtual Conference / Alexandria, VA, USA, January 9 -
               12, 2022},
  pages     = {1204--1220},
  publisher = {{SIAM}},
  year      = {2022},
  url       = {https://doi.org/10.1137/1.9781611977073.50},
  doi       = {10.1137/1.9781611977073.50},
  bibsource = {dblp computer science bibliography, https://dblp.org}
}

@Article{DRS05,
  author	= {Dinur, Irit and Regev, Oded and Smyth, Clifford},
  journal	= {Combinatorica},
  month		= sep,
  number	= {5},
  pages		= {519--535},
  title		= {The Hardness of 3-Uniform Hypergraph Coloring},
  volume	= {25},
  year		= {2005},
  doi		= {10.1007/s00493-005-0032-4},
  issn		= {1439-6912}
}

@article{BartoBKO21,
  author    = {Libor Barto and
               Jakub Bul{\'{\i}}n and
               Andrei A. Krokhin and
               Jakub Oprsal},
  title     = {Algebraic Approach to Promise Constraint Satisfaction},
  journal   = {J. {ACM}},
  volume    = {68},
  number    = {4},
  pages     = {28:1--28:66},
  year      = {2021},
  url       = {https://doi.org/10.1145/3457606},
  doi       = {10.1145/3457606},
  eprinttype = {arXiv},
  eprint = {1811.00970}
}

@article{BrandtsWZ21,
  author    = {Alex Brandts and
               Marcin Wrochna and
               Stanislav Zivn{\'{y}}},
  title     = {The Complexity of Promise {SAT} on Non-{B}oolean Domains},
  journal   = {{ACM} Trans. Comput. Theory},
  volume    = {13},
  number    = {4},
  pages     = {26:1--26:20},
  year      = {2021},
  url       = {https://doi.org/10.1145/3470867},
  doi       = {10.1145/3470867},
  eprinttype	= {arXiv},
  eprint	= {1911.09065},
  
}

@article{Lovasz78,
	author    = {L{\'{a}}szl{\'{o}} Lov{\'{a}}sz},
	title     = {Kneser's Conjecture, Chromatic Number, and Homotopy},
	journal   = {J. Comb. Theory, Ser. {A}},
	volume    = {25},
	number    = {3},
	pages     = {319--324},
	year      = {1978},
	url       = {http://dx.doi.org/10.1016/0097-3165(78)90022-5},
	doi       = {10.1016/0097-3165(78)90022-5},
	bibsource = {dblp computer science bibliography, http://dblp.org}
}

@article{SimonyiZ10,
	author    = {G{\'{a}}bor Simonyi and
	Ambrus Zsb{\'{a}}n},
	title     = {On topological relaxations of chromatic conjectures},
	journal   = {Eur. J. Comb.},
	volume    = {31},
	number    = {8},
	pages     = {2110--2119},
	year      = {2010},
	url       = {http://dx.doi.org/10.1016/j.ejc.2010.06.001},
	doi       = {10.1016/j.ejc.2010.06.001},
	bibsource = {dblp computer science bibliography, http://dblp.org}
}

@inproceedings{AustrinBP20,
  author    = {Per Austrin and
               Amey Bhangale and
               Aditya Potukuchi},
  editor    = {Shuchi Chawla},
  title     = {Improved Inapproximability of Rainbow Coloring},
  booktitle = {Proceedings of the 2020 {ACM-SIAM} Symposium on Discrete Algorithms,
               {SODA} 2020, Salt Lake City, UT, USA, January 5-8, 2020},
  pages     = {1479--1495},
  publisher = {{SIAM}},
  year      = {2020},
  url       = {https://doi.org/10.1137/1.9781611975994.90},
  doi       = {10.1137/1.9781611975994.90},
  bibsource = {dblp computer science bibliography, https://dblp.org},
  eprinttype={arXiv},
  eprint={1810.02784}
}

@article{KOWZ,
  author    = {Andrei A. Krokhin and
               Jakub Oprsal and
               Marcin Wrochna and
               Stanislav Zivn{\'{y}}},
  title     = {Topology and adjunction in promise constraint satisfaction},
  journal   = {CoRR},
  volume    = {abs/2003.11351},
  year      = {2020},
  url       = {https://arxiv.org/abs/2003.11351},
  eprinttype = {arXiv},
  eprint    = {2003.11351},
  bibsource = {dblp computer science bibliography, https://dblp.org}
}

\end{document}